\documentclass[11pt]{article}
\usepackage{fullpage}

\usepackage[margin=3.3cm]{geometry}
\usepackage{amsmath}
\usepackage{dsfont}
\usepackage[latin1]{inputenc}
\usepackage[T1]{fontenc}
\usepackage[active]{srcltx}
\usepackage{amsfonts}
\usepackage{amssymb}
\usepackage{natbib}
\usepackage{graphicx}
\usepackage{color}
\usepackage{bm}
\usepackage{bbm}
\usepackage{enumerate}
\usepackage{ulem}

\newenvironment{proof}{\noindent {\bf Proof }}{\hfill $\bullet$ \vspace{0.25cm}}

\def\E{{\mathbb E}}

\newtheorem{theo}{Theorem}
\newtheorem{prop}{\indent Proposition}

\newtheorem{defin}{\indent Definition}

\newcommand{\su}{\section}
\newcommand{\ssu}{\subsection}
\newcommand{\sssu}{\subsubsection}
\newcommand{\ben}{\begin{enumerate}}
\newcommand{\een}{\end{enumerate}}
\newcommand{\beq}{\begin{equation}}
\newcommand{\eeq}{\end{equation}}
\newcommand{\baR}{\begin{array}}
\newcommand{\eaR}{\end{array}}
\newcommand{\bit}{\begin{itemize}}
\newcommand{\eit}{\end{itemize}}

\newcommand{\Set}[1]{\left\{\, #1 \, \right\}}
\newcommand{\setZ}{\mathbbm{Z}}

\newcommand{\setR}{\mathbbm{R}}
\newcommand{\pare}[1]{\left(#1 \, \right)}

\newcommand{\bra}[1]{\left[#1 \, \right]}
\newcommand{\deq}{\stackrel {\rm def}{=}}
\newcommand\seq[3]{{#1}_{#2}^{#3}}
\newcommand\bloc[2]{{\omega}_{#1}^{#2}}
\newcommand{\Prob}[1]{\mathds{P}\left[\, #1 \, \right]}
\newcommand{\Probc}[2]{\mathds{P}\bra{#1 \, \left| \, #2 \right.}}
\newcommand{\Probcb}[2]{\mathds{P}^{(b)}\bra{#1 \, \left| \, #2 \right.}}
\newcommand{\Probb}[1]{\mathds{P}^{(b)}\left[\, #1 \, \right]}

\newcommand{\om}{\varpi}
\newcommand\blom[2]{{\om}_{#1}^{#2}}

\newcommand\cA{{\mathcal A}}

\begin{document}

\title{On the mathematical consequences of binning spike trains}
\author{B. Cessac\thanks{Biovision team INRIA, Sophia Antipolis, France. \newline \indent INRIA, 2004 Route des Lucioles, 
06902 Sophia-Antipolis, France. \newline \indent   email: bruno.cessac@inria.fr }, A. Le Ny\thanks{Laboratoire LAMA, UMR CNRS 8050, Cr\'eteil, France. \newline  Universit\'e Paris Est Cr\'eteil (UPEC), 91 Avenue du G\'en\'eral de Gaulle, 94010 Cr\'eteil cedex, France. \newline \indent email: arnaud.le-ny@u-pec.fr}, E. L\"ocherbach\thanks{Laboratoire AGM, UMR CNRS 8088, Cergy-Pontoise, France. \newline Universit\'e de Cergy-Pontoise, 2, avenue Adolphe Chauvin, 95302 Cergy-Pontoise Cedex, France.
\newline \indent email: eva.loecherbach@u-cergy.fr}}
\maketitle

\begin{abstract}
We initiate a mathematical analysis of  hidden effects induced by binning spike trains of neurons. Assuming that the original spike train has been generated by a discrete Markov process, we show that binning generates a stochastic process which is \textit{not Markov} any more, but is instead a Variable Length Markov Chain (VLMC) with unbounded memory. We also show that the law of the binned raster is a Gibbs measure in the DLR (Dobrushin-Lanford-Ruelle) sense coined in mathematical statistical mechanics. This allows the derivation of several important consequences on statistical properties of binned spike trains. In particular, we introduce the DLR framework as a natural setting to mathematically formalize anticipation, i.e. to tell "how good" our nervous system is at making predictions. In a probabilistic sense, this corresponds to condition a
process by its future and we discuss how binning may affect our conclusions on this ability. We finally comment what could be the consequences of binning in the detection of spurious
phase transitions or in the detection of wrong evidences of criticality.

\end{abstract}

\su{Introduction} \label{sec:introduction}

The development of multi-electrode arrays (MEA) technology offers an efficient way to record the spiking activity of populations of neurons, in the retina or in the cortex. Currently, up to 4096 neurons can be recorded simultaneously \cite{ferrea-etal:12}. This provides new insights to better understand how a population of neurons encodes information. 

The analysis of MEA data requires however preliminary specific treatments such as spike sorting, which allows to distinguish spikes coming from a specific neuron from the electric variations of potential recorded from nearby electrodes \cite{delescluse-pouzat:06,einevoll-franke-etal:2012,marre-amodei-etal:12}. Once spikes have been sorted, and as spikes are sparse with a spike time subject to some indeterminacy, a usual strategy consists of binning data. That is, one defines first a time window of $\sim 5-20$ ms (binning window), larger than the typical duration of a spike ($\sim 1-2$ ms). The whole spike train is then divided into contiguous, non overlapping such windows and for each neuron a binary variable is defined: it takes value $0$ if the neuron has not spiked in the binning window, and it is $1$ if the neuron has spiked at least once; that is, windows containing $1,2$ or more spikes are all given the value $1$.

Spike sorting, as well as binning, are operations which can have a strong impact on raw data, especially on spike train statistics. In this paper, we concentrate on the mathematical consequences of binning. Assuming that the original spike train has been generated by a  Markov chain (i.e.\ a process with finite memory), we show that binning generates a process losing the Markov property. Instead, the stochastic process describing this raster is a Variable Length Markov Chain (VLMC) with unbounded memory. We will discuss in Section 4 that in some cases such a mechanism is expected to generate long range space and time correlations which might be misunderstood as fallacious evidence of phase transition or criticality. In our situation however, no phase transition and no phenomenon of criticality arises. 


Anticipation plays a key role in the nervous system. A natural question is how "good" our nervous system is at making predictions. To assess the efficiency of prediction, a possible strategy is to measure the information that neurons carry about the future of sensory experiences. In this spirit, illuminating experiments have been done, in the retina, by S. Palmer {\em et al.} \cite{palmer-marre-etal:15}, who conclude that groups of neurons in the retina are indeed close to maximally efficient at separating predictive information from the non-predictive background. From a probabilistic point of view, such prediction amounts to condition a process by its future. In the context of mathematical statistical mechanics, this corresponds to the Dobrushin-Lanford-Ruelle (DLR) approach to rigorously define Gibbs measures and phase transitions \cite{Geo}. DLR measures are measures consistent with regular systems of two-sided conditionings (i.e.\ conditionings w.r.t.\ the outside of finite sets), which corresponds to conditioning w.r.t.\ future and past in dimension one. 

To see whether binning affects the capacity of prediction, we provide a first step towards this direction by proving that the Gibbs property in the DLR sense is preserved in our case. This means, loosely speaking, that the law of the binned raster behaves well in terms of capacity of prediction. We also discuss how one might get other situations where the Gibbs property is lost and what would be the consequences in the detection of spurious phase transitions or in the detection of wrong evidences of criticality.
 \\

The paper is organized as follows. In Section \ref{sec:Sketch} we give the main ideas explaining the effects announced above. Section \ref{sec:Maths} provides a rigorous setting for these statement. Finally, Section \ref{sec:conclusion} is devoted to a discussion on the effect of binning in the context of MEA analysis. All mathematical proofs are given in the Appendix. 
\section{Qualitative description} \label{sec:Sketch}

\ssu{Definitions} \label{sec:def}

\sssu{Spike trains} \label{sec:spike_trains}

We consider the joint activity of $N$ neurons, characterized by the emission of action potentials ("spikes"). The membrane potentials of neurons evolve according to known biophysical mechanisms \cite{dayan-abbott:01,gerstner-kistler:02,ermentrout-terman:10}. Here, we consider that all what we are able to measure is spiking activity, e.g.\ via Multi-Electrode Arrays measurement followed by spike sorting. 

We  also assume that the spiking activity has been recorded at a time scale $\delta$ which is sufficiently small so that a neuron can at most fire one spike within a time window of size $\delta$ (we can set $\delta=1$ without loss of generality). This provides a time discretization labeled with an integer time $n$. Each neuron's activity is then characterized  by a binary variable $\omega_k(n)=1$ if  neuron $k$ fires at time $n$ and $\omega_k(n)=0$ otherwise. 

The state of the entire network at time $n$ is thus described by a vector  $\omega(n) \deq \bra{\omega_k(n)}_{k=1}^{N}$.
A spike block $\bloc{m}{n}$, $n \geq m$, is the sequence of vectors $\omega(m), \omega(m+1), \dots, \omega(n)$; blocks will be denoted by $\bloc{m}{n}$. The "time-range" (or "range") of a block $\bloc{m}{n}$ is given by $n-m+1$ which is the number of time steps needed to go from $m$ to $n$. A "raster" (or "spike train") is a block $\bloc{n_0}{T}$ where $n_0$ is the initial time of the experiment and $T$ the final time. For convenience we shall often consider $n_0 \to -\infty$ and $T \to +\infty$, i.e.\ a bi-infinite raster. Thus, time index runs over $\setZ$. For simplicity,  we shall also use the notation $\omega$ for a raster.

\sssu{Markov chain}\label{sec:MarkovChain}

We consider here the  simple case where the spike train is a realization of a homogeneous Markov chain 
with memory $D>0$. The evolution of the chain is characterized by a family of transition probabilities $\Probc{\omega(D)}{\bloc{0}{D-1}}$. "Homogeneous" means that these transition probabilities do not depend on time. To consider the simplest situation we assume that $\Probc{\omega(D)}{\bloc{0}{D-1}}>0$ for all ${ \bloc{0}{D}}$. Thus, the chain is \textit{primitive}: there exists an $n $ such that for any pair of blocs $w'$, $w$ of range $D$ there is a path 
of length $n$ and of positive probability joining $w'$ to $w$.

An immediate consequence of primitivity is the existence and uniqueness of a unique invariant probability (equilibrium state). Additionally, time correlations decay exponentially fast \cite{seneta:06,bremaud}.

\sssu{Binning}\label{sec:Binning}

We fix an integer $\tau > 1$, the "binning window size". To a raster $\omega$
we associate a binned raster $\om$ defined in the following way. We divide $\setZ$ into contiguous
binning windows, $F_m= [m\tau,(m+1)\tau-1] \cap \setZ$, $m \in \setZ$. We construct\footnote{This construction is similar to scaling transformations with non-overlapping blocks performed within the RG framework, such as decimation  Kadanoff or majority-rule transformations. It is known in mathematical statistical mechanics that these transfomations can generate hidden long-range orders due to untypical discontinuities, see e.g. \cite{VEFS, ELN, kadanoff:66,wilson:75} .
} the binned raster $\om(m) = [\om_k(m)]_{k=1}^N$ 
\beq
\om_k (m) = \left\{
\baR{lll}
1, &\quad  \exists n \in F_m,  \omega_k(n)=1;
\\
0, &\quad \forall n \in F_m,   \omega_k(n)=0.
\eaR
\right.
\eeq
Thus $\om_k(m)=1$ if neuron $k$ has spiked at least once in window $F_m,$ and $\om_k(m)=0$ if neuron $k$ has never spiked in window $F_m$. 

\subsection{The consequences of binning} \label{seq:ConsBin}

\subsubsection{The simplest example} \label{seq:SimpleEx}

To start up, consider the case where $N=1$ (one neuron), $D=1$ (the Markov chain has memory one) and $\tau=2 .$ This simple situation already captures the main features of the procedure. 
Thus, we can drop the neuron index $1$ on the variable $\om_1(m)$ and write $\om(m)$. 

As $\tau=2$,
the binned symbol $\om=0$ corresponds to the successive events $(0,0)$ in the initial raster.
On the opposite,
$\om=1$ corresponds either to $(0,1)$, $(1,0)$ or $(1,1)$. Thus, we associate to the symbol $\om=1$ 
three symbols of the initial Markov chain. This operation is called "factorization". As we will show in Section \ref{sec:Maths}, factorization leads in general to a loss of the Markov property with the creation of a {\it memory of variable length}. We illustrate this here. A general mathematical proof is given in the appendix. \\

We note $\mathds{P}$ the probabilities for the initial chain and $\mathds{P}^{(b)}$ the probability for the binned chain.
We want to show first  how the binned chain loses the Markov property.
For this we first  show that 
\begin{equation}\label{eq:NonMarkovSimple}
\mathds{P}^{(b)} [\om(2)=0 | \om(1)=1,\om(0)=1 ] \neq \mathds{P}^{(b)} [ \om(2)=0 | \om(1)=1 ].
\end{equation}

We have
$$
\mathds{P}^{(b)}[\om(2)=0 | \om(1)=1] = 
\frac{\mathds{P}^{(b)}[\om(2)=0,\om(1)=1]}{\mathds{P}^{(b)}[\om(1)=1]}.
$$

\begin{figure}[!htb]
\centering
 \includegraphics[width=5cm,height=4cm]{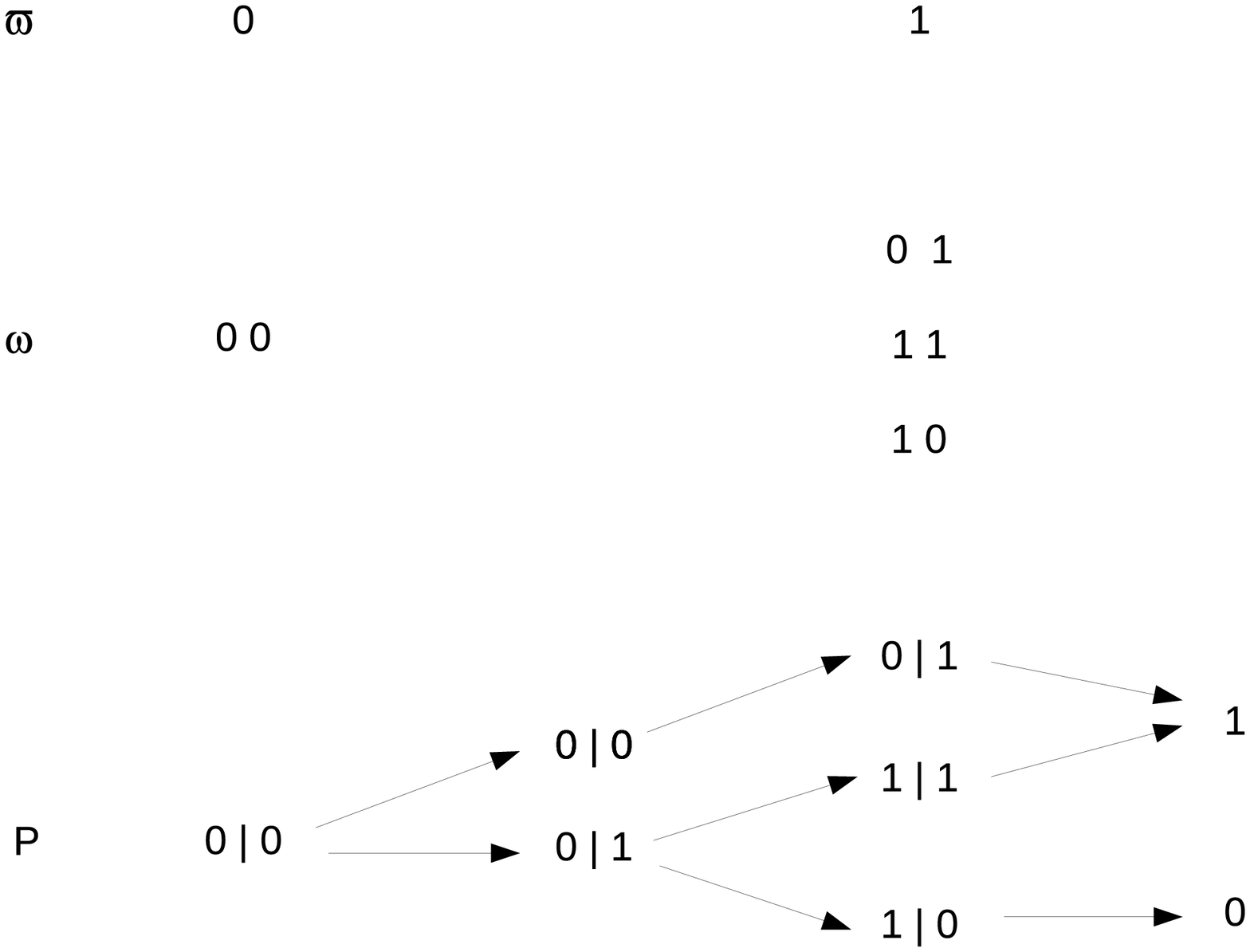}
 \hspace{2cm}
  \includegraphics[width=3cm,height=4cm]{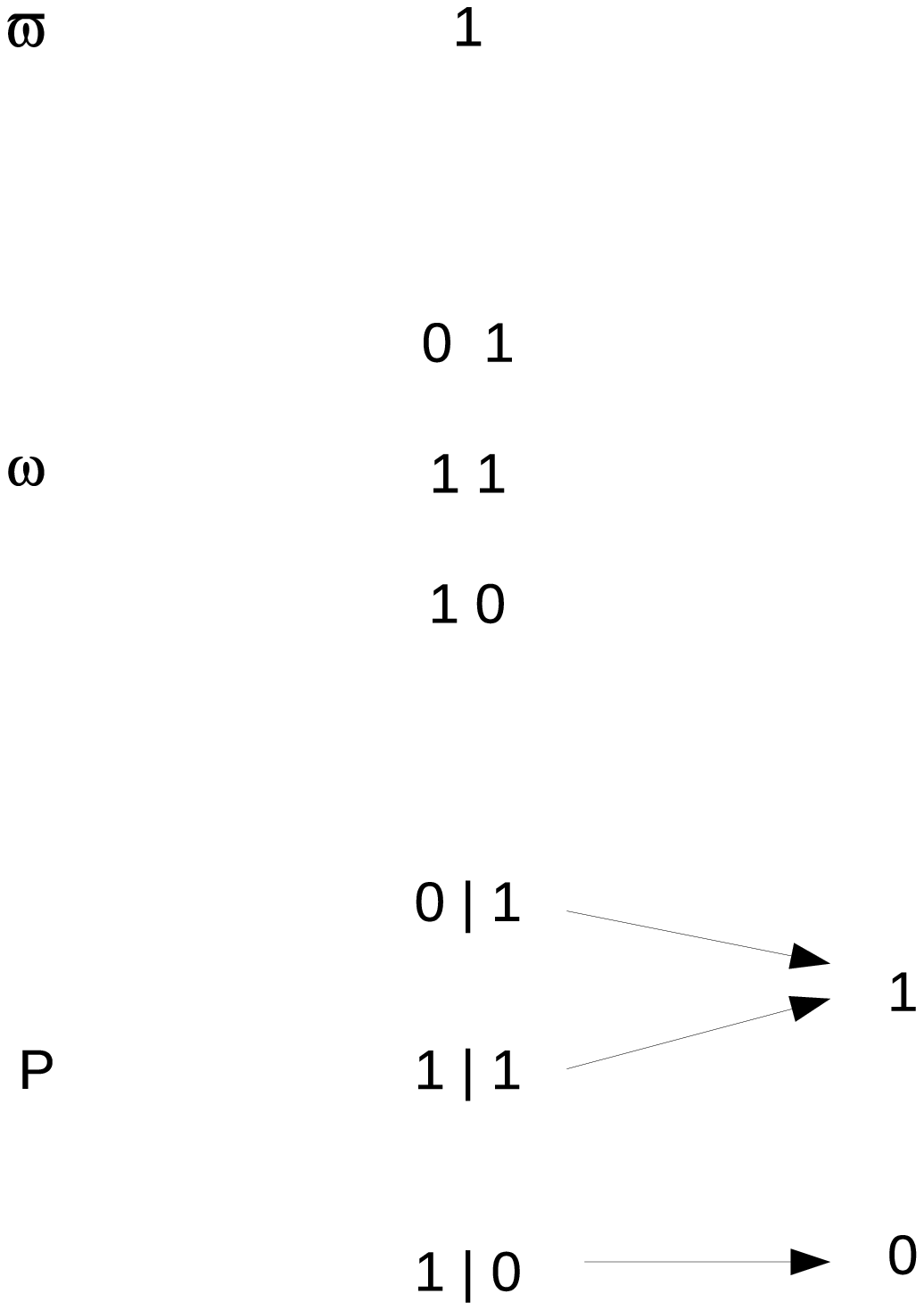}
 \caption{Left: $\Prob{\om(2)=0,\om(1)=1}$. Right: $\Prob{\om(1)=1}$.}
 \label{fig:Transition_Prob_1}
\end{figure}

The complete formulation of $\mathds{P}^{(b)} [\om(2)=0 | \om(1)=1]$ in terms of the transition probabilities $\mathds{P}[\cdot | \cdot]$ of the initial Markov chain is easy using this formula and the Markov property, but the computation is relatively heavy, although simple. Thus, to make computations easier we use a diagrammatic expansion.
An example is given in Fig. \ref{fig:Transition_Prob_1}. \\
In order to compute $\mathds{P}^{(b)}[\om(2)=0,\om(1)=1]$ in terms of the initial chain, one has to consider the symbols appearing on the line "P" of the diagram. Symbol $u | v$
corresponds to $\Probc{u}{v}$, $u$ to $\Prob{u}$, and we read the diagram from the left to the right, going from the present (on the left) to the past (on the right).
One multiplies the probabilities on each path made with arrows, this gives a weight to this path. Finally, one adds all weights to obtain the probability. In this way we obtain (Fig. \ref{fig:Transition_Prob_1} left)
$$
\tiny{
\mathds{P}^{(b)}[\om(2)=0,\om(1)=1] =
\begin{array}{lll}
&\Probc{\omega(5)=0}{\omega(4)=0}\Probc{\omega(4)=0}{\omega(3)=0}\Probc{\omega(3)=0}{\omega(2)=1}\Prob{\omega(2)=1}\\
+ &\Probc{\omega(5)=0}{\omega(4)=0}\Probc{\omega(4)=0}{\omega(3)=1}\Probc{\omega(3)=1}{\omega(2)=1}\Prob{\omega(2)=1}\\
+ &\Probc{\omega(5)=0}{\omega(4)=0}\Probc{\omega(4)=0}{\omega(3)=1}\Probc{\omega(3)=1}{\omega(2)=0}\Prob{\omega(2)=0}
\end{array}
}
$$
and (Fig.\ \ref{fig:Transition_Prob_1} right)
$$
\mathds{P}^{(b)}[\om(1)=1]=
\begin{array}{lll}
&\Probc{\omega(3)=0}{\omega(2)=1}\Prob{\omega(2)=1}\\
+ &\Probc{\omega(3)=1}{\omega(2)=1}\Prob{\omega(2)=1}\\
+ &\Probc{\omega(3)=1}{\omega(2)=0}\Prob{\omega(2)=0}
\end{array}
$$

Finally, $\mathds{P}^{(b)}[\om(2)=0 | \om(1)=1]$ is given by the ratio $\frac{\mbox{Left Diagram}}{\mbox{Right Diagram}}$, where by each "Diagram" we mean the product weight of admissible paths, see below.

As we have implicitly assumed stationarity here, we can in fact drop the time indexes as well, ending up with
\begin{equation}\label{eq:P0s1}
\mbox{\small{$\mathds{P}^{(b)}[0 |1] = \frac{\Probc{0}{0}\Probc{0}{0}\Probc{0}{1}\Prob{1}+ \Probc{0}{0}\Probc{0}{1}\Probc{1}{1}\Prob{1}
+\Probc{0}{0}\Probc{0}{1}\Probc{1}{0}\Prob{0}}{\Probc{0}{1}\Prob{1}
+ \Probc{1}{1}\Prob{1}
+ \Probc{1}{0}\Prob{0}}$}}.
\end{equation}
We shall use this shorthand notation from now on.\\

\begin{figure}[!htb]
\centering
  \includegraphics[width=5cm,height=4cm]{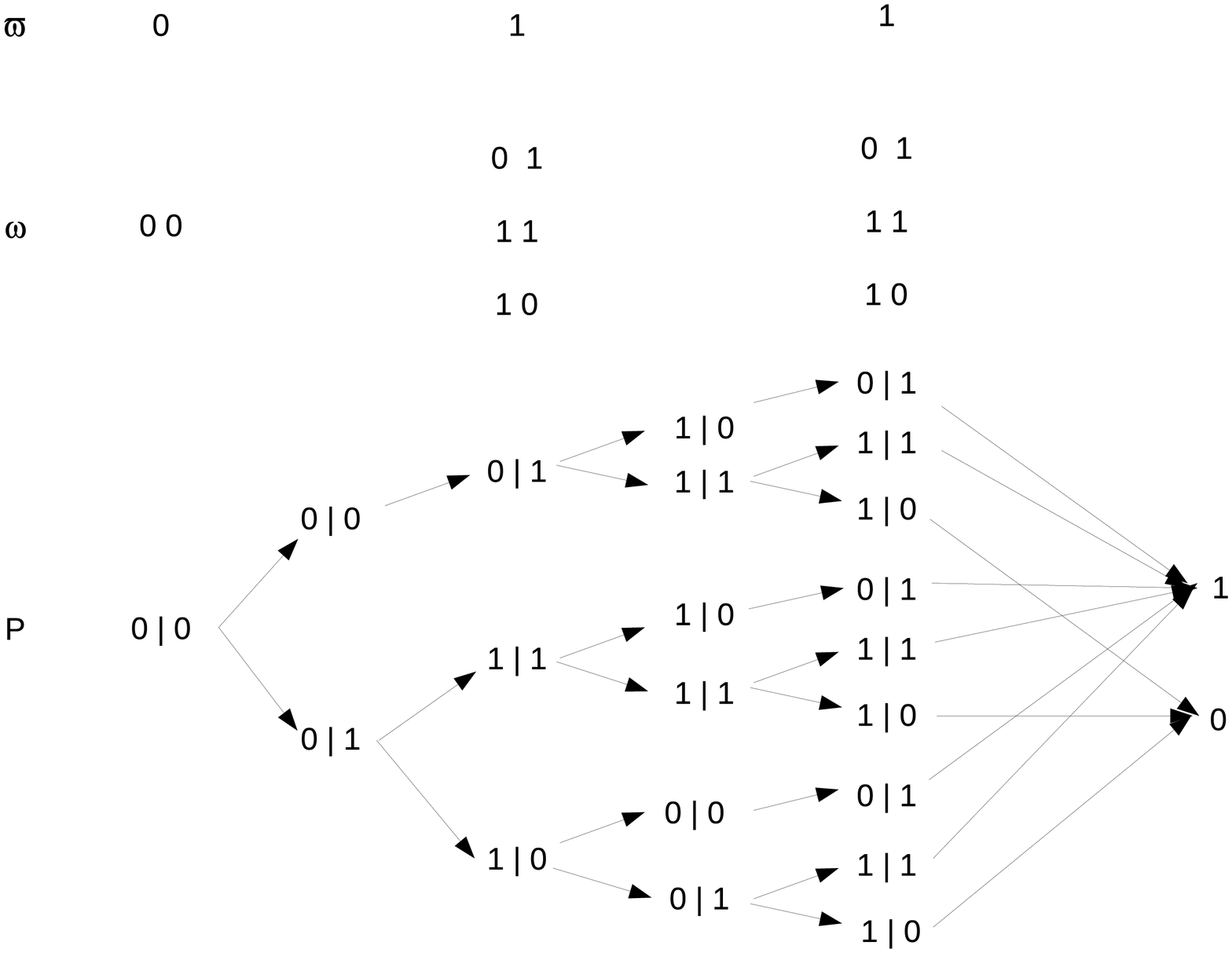}
 \hspace{2cm}
 \includegraphics[width=5cm,height=4cm]{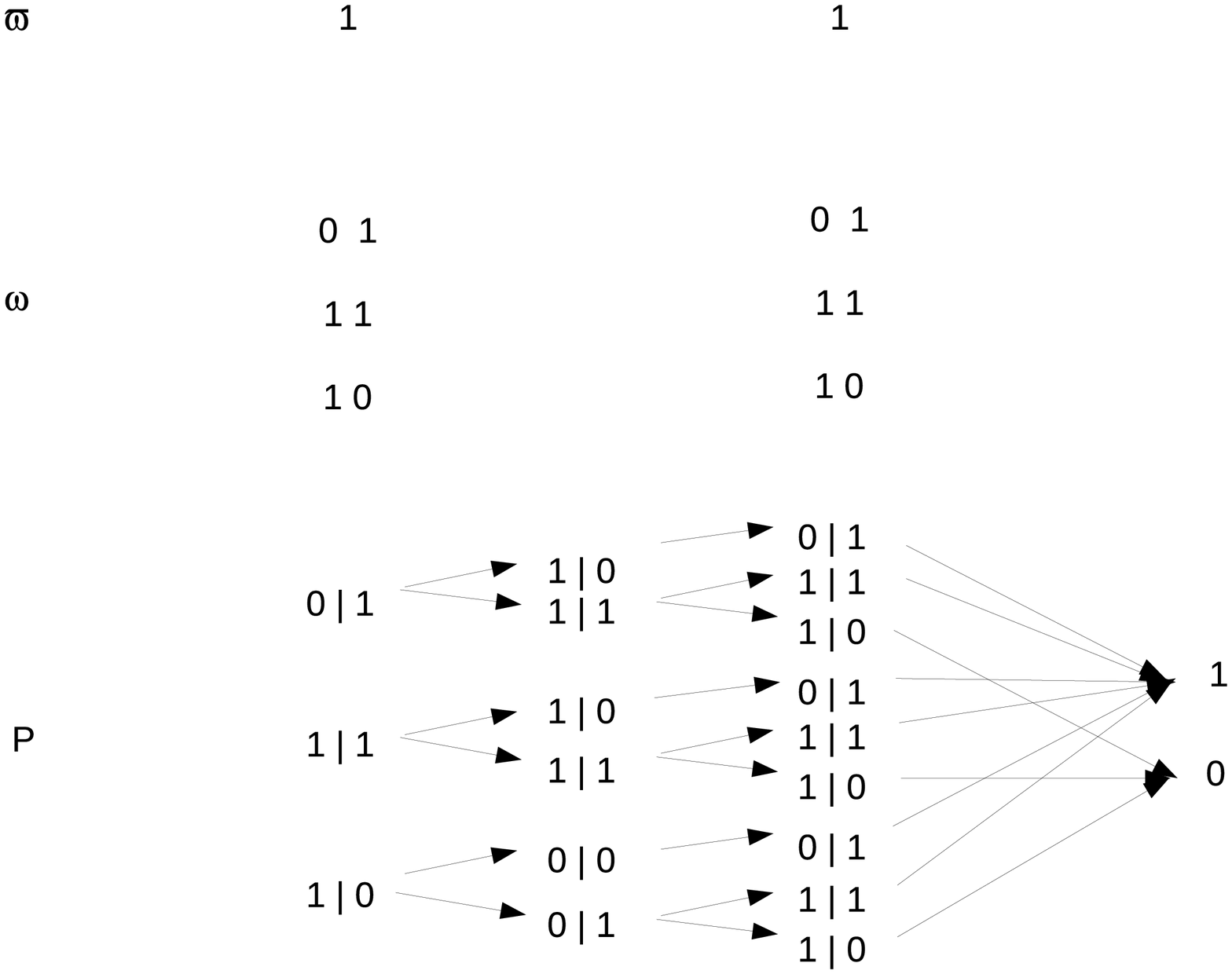}
 \caption{Left: $\Prob{011}$. Right: $\Prob{11}$.}
 \label{fig:Transition_Prob_2}
\end{figure}

From the diagram in Fig.\ \ref{fig:Transition_Prob_2} we obtain 
$\mathds{P}^{(b)}[0|11] = \frac{\mathds{P}^{(b)}[011]}{\mathds{P}^{(b)}[11]}$ as well.


Now, the reason why \eqref{eq:NonMarkovSimple} holds can be readily seen on the graphs. The denominator of left hand side and right hand side are sums corresponding to paths either ending with $\Prob{1}$ or $\Prob{0}$. There is therefore no general way to simplify terms in the numerator and denominator so that the left-hand side equals the right-hand side. \\

\paragraph{Numerical example.}
We instanciate the above discussion by a concrete numerical example. We consider a Markov chain with transitions given by $\Probc{0}{1} = \Probc{1}{0} = \frac34,$ starting from $ \omega (0) = 0 ,$ and propose to calculate 
$  \mathds{P}^{(b)}[\om(2)=0 | \om(1)=1] .$ Applying formula \eqref{eq:P0s1}, we obtain that $\mathds{P}^{(b)}[0 |1] = \frac{5\cdot  3^3 + 3 }{4^2 ( 3^3 + 7) } = \frac{138}{16\cdot 34} \sim 0, 2536.$
On the other hand, we calculate in the same way $\mathds{P}^{(b)}[0 |10 ] = \frac{3^4}{4^2 \cdot 57} \sim 0, 0888 $ which is clearly different from $ \mathds{P}^{(b)}[0 | 1] .$ Therefore, 
$$
\mathds{P}^{(b)} [\om(2)=0 | \om(1)=1,\om(0)=0 ] \neq \mathds{P}^{(b)} [ \om(2)=0 | \om(1)=1 ],
$$
which shows that the binned chain is not Markov of order one any more.\\

Let us come back to our general considerations. The situation is different when the conditioning term contains a ``$0$''. Let us show that
\begin{equation}\label{eq:MarkovSimple0}
\mathds{P}^{(b)}[0 | 101 ] = \mathds{P}^{(b)}[0 | 10].
\end{equation}

\begin{figure}[!htb]
\centering
  \includegraphics[width=5cm,height=4cm]{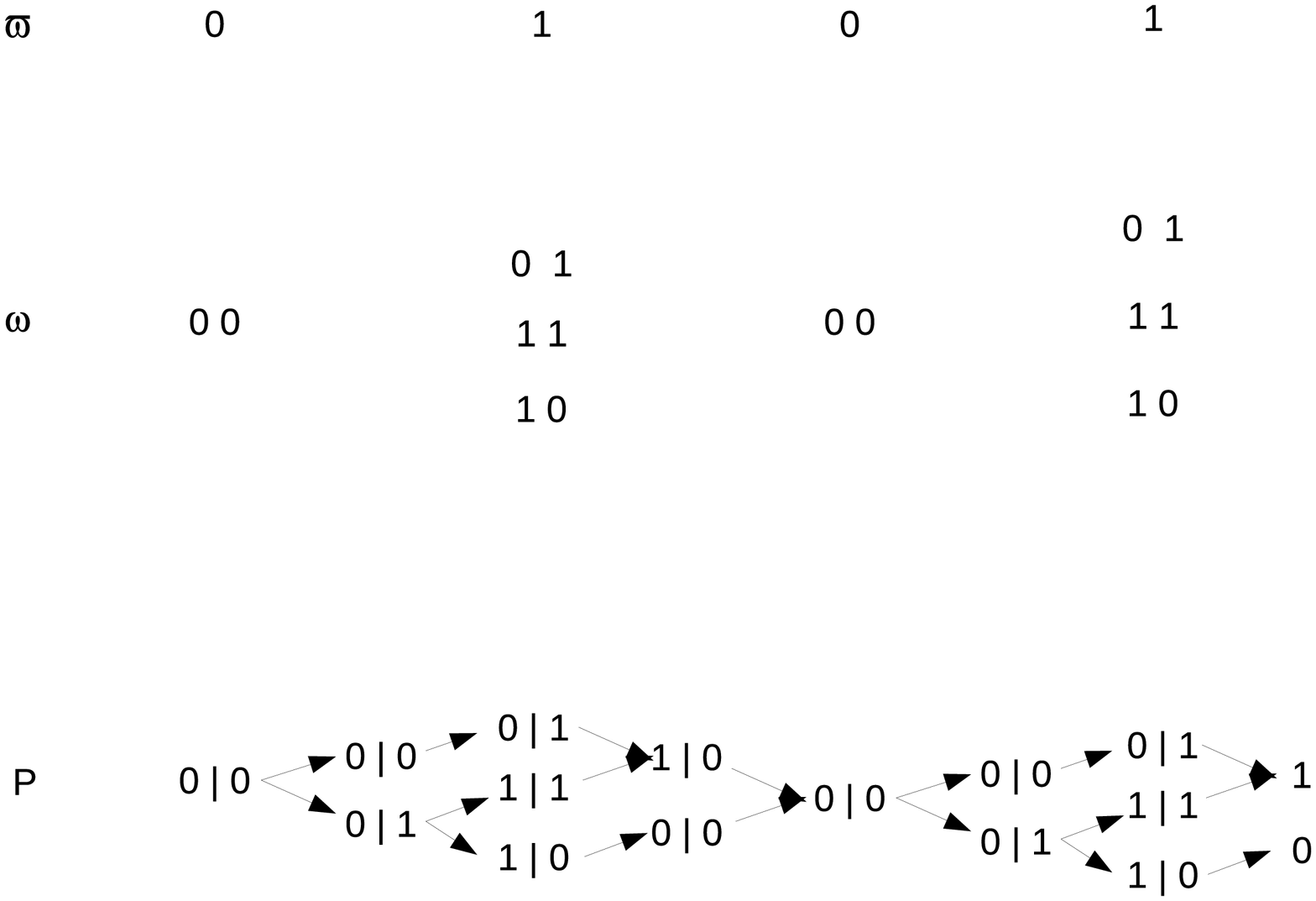}
 \hspace{2cm}
 \includegraphics[width=5cm,height=4cm]{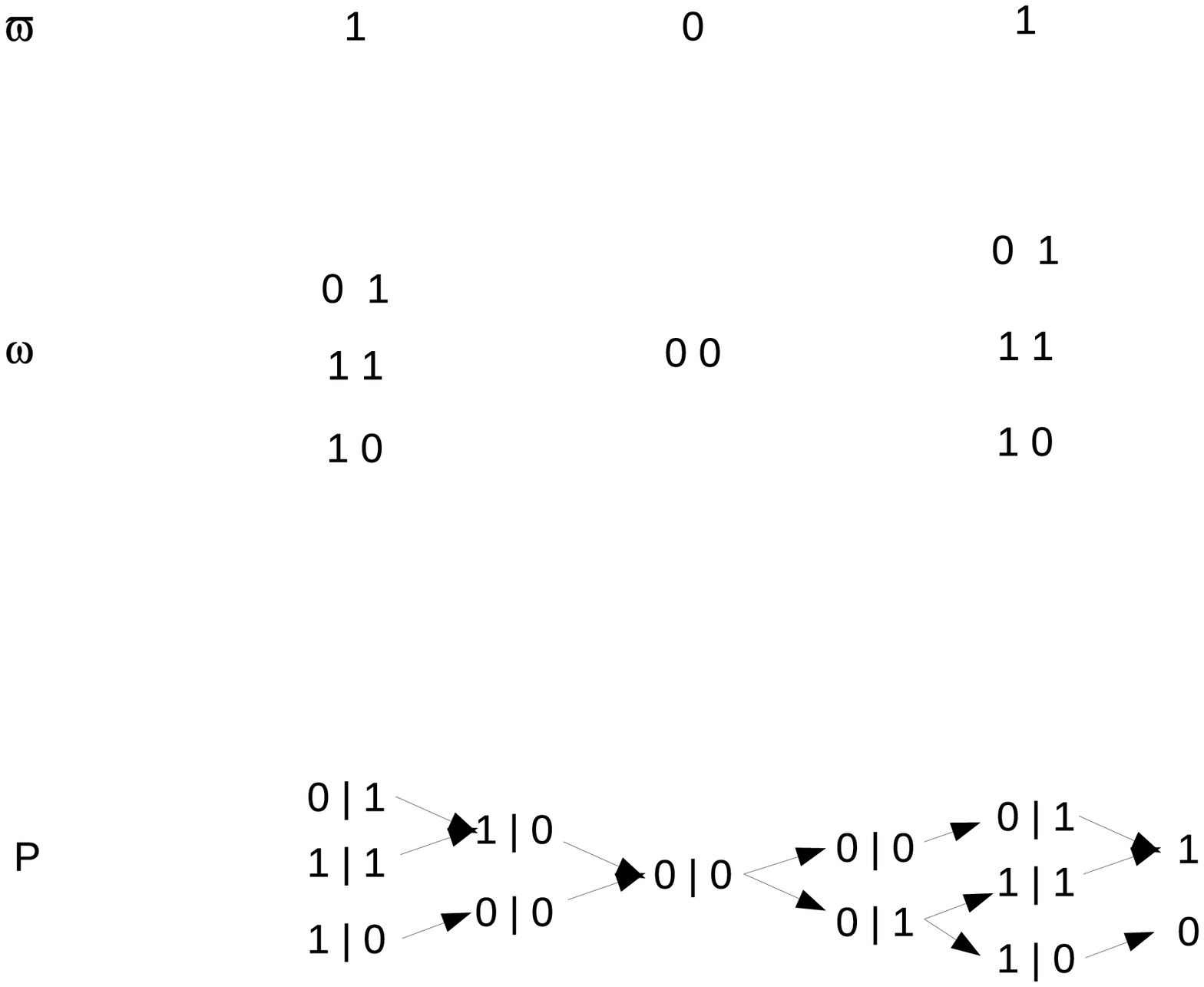}
 \caption{Left: $\Prob{0101}$. Right: $\Prob{101}$.}
 \label{fig:Transition_Prob_3}
\end{figure}

The probability $\mathds{P}^{(b)}[0 | 101] = \frac{\mathds{P}^{(b)}[0101]}{\mathds{P}^{(b)}[101]}$
is represented in Fig. \ref{fig:Transition_Prob_3}.

Applying the rules of our diagrams, we see that $\mathds{P}^{(b)}[0101]$  and $\mathds{P}^{(b)}[101]$ can be factorized
into two subgraphs on the left and on the right of the central symbol $00$. Thus the term on the right of these graphs disappears when computing the ratio $\frac{\mathds{P}^{(b)}[0101]}{\mathds{P}^{(b)}[101]}$ and we end up with $\frac{\mathds{P}^{(b)}[010]}{\mathds{P}^{(b)}[10]}=\mathds{P}^{(b)}[0 | 10]$.\\

We conclude now the above example. For a binned block $\seq{\om}{r}{s}$
we have
\begin{equation}\label{eq:VLMC_simple}
\mathds{P}^{(b)}[\om(s+1) | \seq{\om}{r}{s}]=\mathds{P}^{(b)}[\om(s+1) | \seq{\om}{l}{s}],
\end{equation}
where $l$ is the first occurrence of the symbol $0$ when going from $s$ to $r$ (we set $l = r,$ if $\om$ does not contain the symbol $0$). That is, the binned process is a {\it Variable Length Markov Chain} (VLMC) in the sense of Rissanen \cite{rissanen:83}, where memory goes
back up to the first occurrence of a $0$ in the past (see also next the section where we formalise this idea).

\subsubsection{Generalization} \label{seq:Generalization}

Let us now generalize this statement to $N$ neurons (the rigorous proof is given in Section \ref{sec:Maths} and in the appendix).
Consider first $\tau=2$ and $N=2$. We want to show e.g.\ that
%
$$
\Probcb{\begin{array}{ccc} 0\\0 \end{array}}{\begin{array}{ccc} 1 \, 0\\0 \, 1 \end{array}}
\neq 
\Probcb{\begin{array}{ccc} 0\\0 \end{array}}{\begin{array}{ccc} 1\\0 \end{array}}.
$$

The situation is basically the same as in the previous section.
When computing $\Probb{\begin{array}{ccc} 0 \, 1 \, 0\\0 \, 0 \, 1 \end{array}}$ and $\Probb{\begin{array}{ccc}  1 \, 0\\ 0 \, 1 \end{array}}$, one has to construct a tree 
weighted by the transition probabilities of the initial chain. 
\\

\begin{figure}[!htb]
\centering
  \includegraphics[width=5cm,height=4cm]{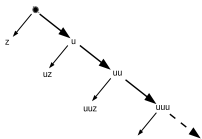}
 \caption{Schematic representation of the binned VLMC.}
 \label{fig:VLMC}
\end{figure}

To further generalize, we classify binned spikes patterns $\om(m)$ in two sets. The first set contains
the unique pattern\footnote{From a general mathematical perspective the fact that $z$ is a block of zeroes plays
no specific role. The same argument works with z being any "end registry". We thank one of the referees for this remark.}, denoted $z$, where all spikes in the window are $0$.  The second
set contains all other patterns (at least one $1$); all elements of this set are denoted by the symbol $u$. Now, \eqref{eq:VLMC_simple} generalizes readily to this case, replacing "$0$" by "$z$" so that $l$ becomes "the first occurrence of the symbol $z$ when going from $s$ to $r$". We provide a schematic representation of this chain in Fig. \ref{fig:VLMC}. 

Let us now discuss the mathematical consequences of the fact that binning leads to a VLMC.

\subsubsection{Consequences} \label{seq:Consequences}

The transformation of a Markov chain into a VLMC via binning has several consequences.
We sketchy present them in this section. The mathematical justifications are given in Section \ref{sec:Maths}.

\begin{enumerate}
\item The memory of the VLMC can extend arbitrarily far in the past. This is the case even if the initial raster is sparse. To obtain the symbol "u" one needs at least one spike in the binned window. In general, the probability of this event increases with $\tau$, the binning window size, and $N$, the number of neurons. 

\item The binned chain has therefore a long range memory, purely induced by binning. 

\item This could induce fallacious long range time correlations as well as  long range space correlations. Let us give a basic example. Assume that neurons are connected on a regular $\setZ^2$ lattice with nearest neighbors interactions which are excitatory. Suppose neuron $i_0$ spikes and triggers a cascade of spikes (avalanche) spreading through the lattice. After $\tau+1$ time steps the avalanche has reached neurons at distance $\tau+1$ from $i_0$. Binning will make those neurons and $i_0$ fire in contiguous time steps. This might create a fallacious causal interaction between them.

\end{enumerate}

So, natural questions arise: "How much does binning impact the estimation of spikes statistics? How far can we mathematically control this impact ?" The next section is devoted
to these questions. 

\section{Mathematical results} \label{sec:Maths}
We write $\cA=\Set{0,1}^N$ for the set of possible vectors with entries $0,1$ in a network of $N$ neurons. The set of spike trains is $\Omega \equiv \cA^\setZ$. Moreover, for any fixed $n \in \setZ,$ we write $\cA_{-\infty}^n $ for all infinite sequences $ ( \omega (k ) )_{ - \infty < k \le n } ,$ and we write $\omega_{-\infty}^n $ for short for such an infinite sequence. 

We introduce the canonical random variable $X_n$ on $\Omega$ defined as a projection by $X_n(\omega)=\omega(n)$ for all $n \in \mathbb{Z}.$ We endow $\Omega $ with the product topology and the associated Borel $\sigma-$algebra ${\cal F} $ which is generated by all projection maps $X_n ,  $ i.e.\ ${\cal F} = \sigma \{ X_n , - \infty < n < \infty  \} .$ Finally we introduce ${\cal F}_{- \infty}^{k  } = \sigma \{ X_n , - \infty < n \le k  \} ,$ the history before time $k.$ Moreover, for any subset $ \Lambda \subset \setZ , $ we introduce ${\cal F}_\Lambda = \sigma \{ X_n , n \in \Lambda \} . $ 

A spike train $\left(\omega(n)\right)_n$ can be seen as a stochastic process defined on the space $(\Omega,\mathcal{F}), $ and its law is entirely determined by a probability law $\mathbb{P}$ on $(\Omega,\mathcal{F}) .$

\subsection{Transition probabilities} \label{sec:TransProb}
In this paper, we want to analyze the effects of binning on inferring the present statistics of spike, given the history, but we also want to consider its effects on anticipation mechanism, where one conditions the present statistics by possible futures.

In a mathematical setting, this leads us to consider  "one-sided" or "two-sided" dependencies. By "one-sided", we mean that the conditioning depends only on one side of the evolution -- the past -- and not on the other side -- the future. By "two-sided", one means that the conditioning is prescribed outside finite sets from both "sides". In dimension one this amounts to considering a prescribed future.  The two notions are not mathematically equivalent. In the simple case of range one dependencies, they are designed by {\it Local Markov} versus {\it Global Markov} properties, see \cite{Foll, Gold} or \cite{FP} in higher dimension. In this paper we are considering a one dimensional situation, where the dimension is time. But, more generally, considering one-sided or two-sided conditionings in larger dimensions is precisely the topic of rigorous mathematical statistical  mechanics, see \cite{VEFS,Geo}.  These notions are also related to equilibrium states and Gibbs properties. We warn the interested reader that the vocabulary can change depending on  the point of view which is adopted. Indeed, slight differences exist between the approaches coming from probability theory \cite{Geo}, mathematical statistical mechanics \cite{Dobrushin} or dynamical systems/ergodic theory \cite{bowen:98}. See also \cite{Robertoetal} for a discussion on different notions.

For the sake of clarity and to avoid usual confusions, we precise here what we mean by Markov chains or fields, and which (different) Gibbs measures might be considered.\\

In the one-sided setting, we focus first on the probabilistic framework and consider stochastic processes $(X_n)_{n \in \mathbb{Z}}$ defined on $(\Omega, \mathcal{F})$. We follow \cite{Robertoetal} and introduce systems of transition probabilities:

\begin{defin}
A {\bf system of transition probabilities} (or transition kernels) is a family $\{p_n(\cdot | \cdot) : n \in \mathbb{Z}\}$ of functions $p_n : \cA \times \cA_{-\infty}^n \longrightarrow [0,1]$, such that the following conditions hold for all $n \in \mathbb{Z}.$
\begin{enumerate}
\item {\bf Measurability :} For each $a \in \mathcal{A},$ the function $p_n(a | \cdot)$ is ${\cal F}_{-\infty}^{n-1} -$measurable. 
\item {\bf Normalization :} For each $\omega_{-\infty}^{n-1} \in \mathcal{A}_{-\infty}^{n-1}$, 
$$
\sum_{a \in \mathcal{A}} p_n \big(a | \omega_{-\infty}^{n-1} \big)=1 .
$$
\end{enumerate}
\end{defin}
A system of transition probabilities defines the intrinsic dynamics of the process $ (X_n)_n :$ $p_n ( a| \omega_{- \infty }^{n-1} ) $ gives the probability of the event $\{X_n = a \} , $ knowing that $ X^{n-1}_{-\infty} = \omega_{-\infty}^{n-1} .$ In other words, the law $ \mathbb{P}$ of $ (X_n)_{n= - \infty}^{n=  \infty} $ corresponds to the dynamics prescribed by $(p_n)_{n \in \mathbb{Z}}.$ This is formalized in the following definition. 

\begin{defin}
A probability measure $\mathbb{P}$ on $(\Omega,\mathcal{F})$ is {\bf consistent} with a system of transition probabilities $(p_n)_n$ if
\begin{equation}
\mathbb{P} \big[X_n=\omega(n) | X_{-\infty}^{n-1}=\omega_{-\infty}^{n-1} \big] = p_n \big(\omega(n) | \omega_{-\infty}^{n-1} \big) ,
\end{equation}
for all $n \in \setZ$ and for $\mathbb{P}-$almost all $\omega_{-\infty}^{n-1} \in \cA_{-\infty}^{n-1} . $
\end{defin}

Here, $ \Probc {X_n = \omega (n)  }{ X_{-\infty}^{n-1}=\omega_{-\infty}^{n-1}}$ denotes a {\it regular version
\footnote{This technical notion means the following: The transition kernel $p_n \big(\omega(n) | \omega_{-\infty}^{n-1} \big)$ is one possible choice -- within the $L^2-$equivalence class of possible choices -- ensuring that for all $ A \in    {\cal A}_{- \infty}^{n-1 } , $ $ \mathbb{E} ( 1_{\{X_n = \omega (n)\}} 1_A ) = \mathbb{E} ( p_n ( X_n | X_{-\infty}^{n-1} ) 1_A ).$ The notion {\it regular} does not refer to any regularity of the function $ \omega_{-\infty}^{n-1} \mapsto p_n \big(\omega(n) | \omega_{-\infty}^{n-1} \big) $ here, it is rather related to the fact that - very roughly speaking - it is possible to condition on an event of probability $0$, that is, to condition on the event $ \{ X_{k } = \omega (k ) , - \infty < k \le n-1 \} .$} of the conditional probability $\Probc {X_n  =\omega (n)  }{ {\cal F}_{- \infty}^{n-1 }}$} on the event $ \{ X_{k } = \omega (k ) , - \infty < k \le n-1 \} .$

Let us emphasize that \textit{stationarity is not assumed in these definitions and the kernel $p_n$ may depend on $n$ as well}. This means that seasonality can be included in the definition of the dynamics, although in the following we mainly assume stationarity.

Other related notions are those of {\it stochastic chains with memory of variable length} (VLMC, see below and \cite{galves-locherbach:08}), chains with complete connections,  g-measures or chains of infinite order (see e.g.\ \cite{Robertoetal}, \cite{Verbitskiy2011315} and the references cited therein). 

Homogeneous and primitive Markov chains are a particular case of such processes, see Definition \ref{defin:markov} below.  In the following, we assume stationarity so that it is sufficient to consider kernels $p $ defined on $ \cA \times \cA_{-\infty}^{-1} .$
Often, some additional continuity conditions (in the product topology of the discret topology on our alphabet) are required that we recall now. 

\begin{defin}\label{defin:reg}
A transition kernel $p :  \cA \times \cA_{-\infty}^{-1}  \to  [0, 1 ] $ is {\bf continuous}  if for all $ a \in \cA $ and for all $ x_{-\infty}^{-1} , $ $ p(a| x_{-k}^{-1}y_{-\infty}^{-k - 1 } ) $ converges as $k \to \infty, $ for any $y_{- \infty}^{-1} .$ Here,  $x_{-k}^{-1}y_{-\infty}^{-k - 1 } $ denotes the concatenated past given by the left-infinite sequence having element $ y (l) $ for any $ l \le -k - 1 $ and $ x (l) $ for any $ - k \le l \le -1.$ 

The continuity rate $ \beta ( k) \equiv \beta^p ( k) $ of the kernel $p$ is defined by 
$$ \beta (k) = \sup_a \sup_{ x_{-k}^{-1} } \sup_{ y_{-\infty}^{ -1} , z_{-\infty}^{-1} } \big | p( a | x_{-k}^{-1} y_{-\infty}^{-k - 1 } ) - p( a | x_{-k}^{-1} z_{-\infty}^{-k - 1 } ) \big | .$$ 
\end{defin}
In other words, the dependency of the past decays with the distance between present and past, and this decay is described by the rate $ \beta ( k) .$ 

Markov chains are the simplest (non-independent) examples of processes consistent with a continuous kernel, because the conditioning depends on the immediate past only -- more precisely, the immediate future $X_{n+1}$ depends on the present $X_n$ only. By extension, we call Markov chain of order $D$ a stochastic chain having a transition kernel $p$ which depends only of a finite portion, of length $D$, of the past. Such processes are trivial examples of processes having continuous transition kernels. 

\begin{defin}\label{defin:markov}
A system of transition kernels $ p_n : \cA \times \cA_{-\infty}^{n-1} \to [0, 1 ] $ is a system of {\bf Markov kernels of order $D > 0 $} if for all $a \in \cA , \omega \in \Omega $ and for all $n \in \setZ, $ $p_n ( a | \omega_{-\infty}^{n-1} ) $ depends only on $ \omega_{n-D}^{n-1} .$ A  probability measure $\mathbb{P}$ on $(\Omega,\mathcal{F})$ is called {\bf Markov measure of order $D$ } if it is consistent with a system of transition probabilities $(p_n)_n$ which are Markov kernels of order $D.$ 
\end{defin}

In this frame, a stochastic chain $(X_n)_{n \in \setZ} $ canonically defined on $(\Omega, \cal F , \mathbb{P}) $ is a {\it Markov chain of order $D$} if and only if $\mathbb{P}$ is a Markov measure of order $D.$

For homogeneous Markov chains of order $D,$ i.e.\ chains where $p_n$ does not depend on $n,$ it suffices to study $ p ( \omega ( 0) | \omega_{-\infty}^{-1} ) .$ Since this transition does only depend on $\omega_{-D}^{-1}, $ we shall write $ p ( \omega ( 0) | \omega_{-D}^{-1} ) $ instead of $ p ( \omega ( 0) | \omega_{-\infty}^{-1} )$ in this case.

\subsection{Modeling spike trains as Markov chains and binning}\label{sec:BinningBis}
We model spike trains as realizations of a homogeneous and primitive Markov chain of order $D>0$ having a transition kernel $ p ( \omega ( 0) | \omega_{-D}^{-1}) ,$  where we suppose that $ p ( \omega ( 0) | \omega_{-D}^{-1}) >0$ for all $\omega_{-D}^0 .$ We write $\mathbb{P}$ for the unique Markov measure on $(\Omega, \cal F) $ which is consistent with this family of transition probabilities, where consistency has to be understood in the sense of Kolmogorov's consistency of marginals \cite{bremaud}.

As in Section \ref{sec:Binning}, we associate a binned raster $(\om (n) )_n $ to a raster $(\omega(n) )_n.$ Moreover, we extend the notion of binning to binned blocks and write for any block $ a_l^m \in {\cal A}_l^m,$ 
$\blom{l}{m} = \seq{a}{l}{m}$ if all $\om(r)=a(r) , \,  r=l , \dots , m .$ 
We define a map $\pi: {\cA}_0^{ \tau -1 } \to \cA ,$ $\pi = [\pi_k ]_{k=1}^N ,$ called \textit{aggregation map}, by
\begin{equation}\label{eq:pi}
 \pi_k ( \omega_0^{\tau -1}  ) = \left\{ 
\begin{array}{ll}
0, & \omega_k ( n ) = 0 \mbox{ for all }  0 \le n \le \tau - 1; \\
1, & \mbox{ else}. 
\end{array}
\right.
\end{equation}
$\pi$ is extended in a canonical way to finite sequences $ \omega_{l\tau }^{ (m+1) \tau -1} .$ As in Section \ref{seq:Generalization}, we write $z$ for the element $[0]_{k=1}^N \in \cA.$ Thus, the set
$\pi^{-1}\pare{\seq{z}{l}{m}}$ is the set of blocks $\bloc{l\tau }{(m+1)\tau - 1}$, in the original raster,  such that each window $F_r$, $r=l  ,\ldots, m,$ contains only $0'$s. Finally, we denote by $\mathbb{P}^{(b)} $ the law of the binned raster, i.e.\ $\mathbb{P}^{(b)} = {\cal L} ( ( \pi ( X_{n \tau}^{(n+1) \tau - 1} )_{n \in \setZ} )| \mathbb{P} ) .$ We write $\E^{(b)}$ for the associated expectation whereas $ \E $ denotes expectation with respect to the original measure $\mathbb{P}.$

\subsection{The binned raster is a Variable length memory chain (VLMC)}\label{sec:ConsBinning}

An immediate consequence of binning is that the resulting chain is not Markov any more, as we have argued in Section \ref{seq:ConsBin}. Indeed, as we will show in this section, now rigorously and in a more general set-up, it is a chain of infinite memory having variable length memory. Stochastic chains with memory of variable length constitute an
  interesting family of stochastic chains of infinite order on a
  finite alphabet. The idea is that for each infinite past,  a finite
  part of the past is enough to predict the
  next symbol,  but the length of this past varies and can be arbitrarily long.  These models were first introduced in the information
  theory literature by Rissanen \cite{rissanen:83} as a universal tool to perform
  data compression. For more details, we refer the reader to \cite{galves-locherbach:08} for a survey of the subject. 

In our framework, the variable length memory structure is given as follows. Write $ \cA_{-\infty}^{ - 1  }$ for the set of all infinite pasts and define for any $ x_{-\infty}^{-1} \in \cA_{-\infty}^{ - 1  },$
$$ l(x_{-\infty }^{-1} ) = \inf \, \{ m : x (-m)  = z \}$$  where we recall that $z$ denotes the "null" configuration for which no neuron has spiked.  
Here, by convention, $\inf \emptyset = - \infty$. Thus, $l(x_{-\infty }^{-1} )$ is the
first index, in the binned raster and back to the past starting from $-1$, where the symbol $z$ is met (i.e. the corresponding block contains no spike). 

The following proposition shows that the memory of the chain is precisely $l(x_{-\infty }^{-1} )$ , as anticipated in Section \ref{seq:Generalization}. Thus, the length of the memory depends on the spike sequence.

\begin{prop}\label{prop:1}
Suppose that $ \tau \geq D .$ Then 
for any infinite past $x_{-\infty}^{-1} $ belonging to $  \cA_{-\infty}^{-1} $ and any symbol $a \in \cA , $
\begin{equation}\label{eq:real}
 \mathbb{P}^{(b)} \left[ X(0) =a |X_{-\infty}^{-1}=x_{-\infty}^{-1}\right] = 
 \mathbb{P}^{(b)} \left[ X(0) =a | X_{- l(x_{-\infty}^{-1})}^{-1}=x_{-  l(x_{-\infty}^{-1})}^{-1} \right]\, .\, 
\end{equation}
\end{prop} 

Here, $\mathbb{P}^{(b)} \left[ X(0) =a |X_{-\infty}^{-1}=x_{-\infty}^{-1}\right]$ denotes a version of the conditional probability $\mathbb{P}^{(b)} \left[ X(0) =a | {\cal F}_{- \infty}^{-1 } \right]  .$ The proof of this proposition is postponed to the appendix. \\

As a consequence, consider the tree $\mathcal{T}$ represented in Figure \ref{fig:VLMC} and defined by  
$$ \mathcal{T} = \{  a_1^k z , k \geq 0 , a(i ) \in \cA, a(i) \neq z, 1 \le i \le k \} ,$$
where $a_1^k z  $ represents the sequence $x_{-k -1}^{-1} $ such that $x (-i)  = a(k-i+1) $ for all $1 \le i \le k $ and $x (-k
  -1)  = z . $ We associate transition probabilities to each leaf $  a_1^k  z $ of the tree via 
$$ p( a |  a_1^k z ) = \mathbb{P}^{(b)} \left[ X(0) =a | X_{- 1} =a_1, X_{-2} = a_2, \ldots , X_{-k } = a_k, X_{-k - 1} =  z  \right] .
$$
The ordered pair $(\mathcal{T} , p) ,$ where 
$$ p = \{ p(.|  a_1^k z   ) ,  k \geq 0  \} $$
is called {\it probabilistic context tree on } $\{0, 1 \}^N .$ It defines entirely the evolution of the binned chain $(X_n)_{n \in \setZ}  $ on $(\Omega, {\cal F} , \mathbb{P}^{(b)}) $ by \eqref{eq:real}.

Therefore, the binning procedure gives rise to a process which is a {\it Variable Length Markov Chain} (VLMC) where the memory extends to the last symbol $z$ encountered in the past. Such a process is not Markov anymore, and we cannot bound a priori the memory depth of the chain. Indeed, this memory can go quite far back into the past. However, 
$$ {\mathbb P }^{(b)} ( \exists \mbox{ infinitely many binning windows } F_m \mbox{ such that } X_m  \equiv z ) = 1 ,$$
since the original chain is primitive. Thus, with probability $1,$ the initial raster gives rise to a binned raster whose transition probabilities have memory depth which is finite (but not fixed, since the memory depends on the realization of the blocks).

\ssu{Continuity properties of the binned transition operator} \label{sec:StillGibbs}
The transition kernel $p^{(b)}$ of the binned chain $(\om ( n) )_n$ is defined by:
\begin{eqnarray}\label{eq:transitionp}
p^{(b)} : \cA \times \cA_{ - \infty}^{-1} & \to & [0, 1 ] \nonumber \\
(a, x_{-\infty}^{-1}) & \mapsto & p^{(b)} [ a | x_{-\infty}^{-1}] :=  \mathbb{P}^{(b)} \left[ X_0   =a | X_{-\infty}^{-1}=x_{-\infty}^{-1} \right]  .
\end{eqnarray} 
By Proposition \ref{prop:1},  $p^{(b)} [ a| x_{-\infty}^{-1} ]$ depends only on $ x_{-\ell }^{-1} ,$ if $ l( x_{-\infty}^{-1} ) = \ell .$ But since $ l( x_{-\infty}^{-1} )$ is not bounded, $p^{(b)}$  is not of bounded memory.

We expect that the influence of past events on the probability $ p^{(b)} ( a| x_{-\infty}^{-1}) $ of $a$ decreases with their distance to this event: the further one goes back to the past, the less the past events influence the present.  
This question is related to the continuity properties of the kernel $p^{(b)}$.  In general, this effect can decay either fast (e.g.\ exponentially) or slowly (e.g.\ sub-exponentially or algebraically). The consequences are quite different. In our situation, this decay is exponential as shows the following proposition.


\begin{prop}\label{prop:2}[Theorem 3.1 of \cite{chazottes-ugalde:11}]
The transition kernel $p^{(b)}$ of the binned chain is continuous and there exists a constant $\alpha > 0 $, depending on $N$, such that  
$$ \beta^{p^{(b)}}  ( k ) = O ( e^{- \alpha {k}} ) \mbox{ as }  k \to \infty .$$
\end{prop}

As a consequence of the above proposition, the transition operator of the binned chain is of infinite memory, but it is continuous and it has furthermore an exponential decay of  the continuity rate. Note however that the decay rate, $\alpha$, will in general depend on the number of neurons (see Section \ref{sec:conclusion}).

\ssu{Does binning affect anticipation ? } \label{sec:Anticipation}
In the context of VLMC's or more generally of chains of infinite order, exponential continuity is enough to ensure the existence and uniqueness of a probability
measure consistent with the system of transition probabilities (see Section \ref{sec:TransProb} and/or \cite{fernandez-maillard:05}). This is a sought property as it means that there is
a unique invariant probability for the chain. However   
Proposition \ref{prop:2} states only the continuity of the one-sided transition probabilities obtained by conditioning upon the past. If we want to consider the effects of binning on anticipation, we might also want to consider conditioning on the future. In mathematical terms, this amounts of considering the Gibbs property in the DLR sense of mathematical statistical mechanics. In general, continuity with respect to the past does not imply that the law of the binned chain 
is also Gibbs in the DLR sense  (\cite{Geo, Dobrushin}, see also \cite{Robertoetal} where an example is exhibited). Therefore, it is not a priori clear that the law of the binned raster is Gibbs in the DLR sense.



In the present section, we show that the law of the binned raster is a Gibbs measure in the DLR sense, i.e.\ that, roughly speaking, it behaves well when also conditioning with respect to the future. Here, {\it to behave well} means that  the binned chain possesses the same good anticipation properties as the non-binned original chain and that a law of large numbers holds as well as good mixing properties. The main reason why this is so here  is that the binned chain has very good regularity properties, i.e.\ the continuity rate is exponential \footnote{Note that  this condition is not necessary.}as shown in Proposition \ref{prop:2}.

Let us start be recalling the following definitions of Gibbs measures in the DLR sense from mathematical statistical mechanics (see e.g. \cite{Robertoetal}).

\begin{defin}
A {\bf specification} is a family of transition kernels $ \gamma= \{\gamma_{\Lambda}\}_{\Lambda \subset  \setZ , |\Lambda | < \infty } , $ $\gamma_\Lambda : {\cal F} \times \Omega \to [0, 1 ] , $ on $(\Omega, {\cal F} )$ such that  
\begin{itemize}
\item[(a)] For each $\Lambda \subset \setZ , | \Lambda | < \infty $ and each $ B \in {\cal F}$, the 
function $ \gamma_{\Lambda}(B | \cdot\,)$ is ${\cal F}_{\Lambda^c}-$measurable. 

\item[(b)] For each $\Lambda \subset \setZ , | \Lambda | < \infty $ and each $ B \in
  {\cal F}_{\Lambda^c}$, $ \gamma_{\Lambda}(B | \omega)= 1_{B}(\omega).$

\item[(c)] For any pair of regions $ \Lambda$ and $\Delta$, with
  $\Lambda \subset \Delta \subset \setZ , | \Delta | < \infty ,$ and any $ B \in {\cal F}$, 
 
\begin{equation}
 \int_\Omega  \gamma_\Lambda( B | \omega ') \gamma_{\Delta}(d\omega ' |  \omega) 
 \;=\; \gamma_{\Delta}(B | \omega)
 \label{eq:3}
\end{equation}
for all $\omega\in\Omega$.
\end{itemize}

\end{defin}

In our frame, we are mainly interested in {\it positive} specifications, i.e.\ $\gamma_\Lambda ( B| \omega ) > 0 $ for all $ B \neq \emptyset, $ for all $ \omega \in \Omega.$ In this case, a specification is uniquely determined by the so-called {\it one-point specification} $ \{  \gamma_{\{i\}} (\cdot|  \omega ), \,i \in \setZ , \omega \in \Omega\},$ see \cite{Geo}.  We write for short $ \gamma_i (  \omega (i )  | \omega ) = \gamma_{\{i\}} ( \{ \omega (i ) \} |  \omega) .$ Intuitively, this is a candidate for the conditional law of $X_i $ conditionally on $ \{ X_n = \omega (n ) , n \neq i \} .$ 

We can now introduce the notion of continuity for specifications as we did before for systems of transition probabilities, see again \cite{Robertoetal}.

\begin{defin}
\begin{itemize}
\item[(a)]
A specification $ \gamma $ is called {\bf continuous} if  for all $i \in \setZ , $ $\gamma_i ( \omega (i ) |  \cdot ) $ is continuous for all $ \omega (i ) \in \cA , $ i.e. 
$$ \sup_{ x , y \in \Omega : x_{-n}^m = y_{-n}^m } | \gamma_i ( x(i) | x ) - \gamma_i ( y(i) | y ) | \to 0 $$
as $ n , m \to \infty .$
\item[(b)] A specification $ \gamma $ is called {\bf strongly non-null} if there exists a constant $c > 0 $ such that for all $\omega (i ) \in \cA , $ $ \gamma_i ( \omega(i ) | \omega ) \geq c > 0 .$ 
\end{itemize} 
\end{defin} 
Thus, being "continuous" means that at the same time the dependency on the past and on the future decays with their distance to the present. Condition (a) without (b) concerns so-called {\it quasilocal specifications} in mathematical statistical mechanics \cite{Geo}.

We now recall the notion of a Gibbs measure in the sense of mathematical statistical mechanics, i.e.\ in the DLR sense. 

\begin{defin}
A shift invariant measure $ \mathbb{P} $ on $(\Omega , {\cal F }) $ is a {\bf Gibbs measure} if it is consistent with a continuous and strongly non-null specification $\gamma , $ i.e.\ for all $ i \in \setZ, $  
$$ \mathbb{P} \left[ X_i = \omega (i ) | X_{- \infty}^{ i-1} = \omega_{-\infty}^{i-1}, X_{i+1}^\infty = \omega_{i+1}^\infty \right] = 
\gamma_i ( \omega (i) | \omega ) $$
for $\mathbb{P}-$almost all $ \omega \in \Omega .$
\end{defin}

Thus, roughly speaking, a Gibbs measure corresponds to  a chain giving weight to every event (non-nullness) where the influence of both past and future on the current state decays with the distance. 

Let us now return to the problem stated in the beginning of this section: Is there a unique probability measure compatible with a given past and a given future ? In mathematical terms, this means : Is the invariant measure $\mathbb{P}^{(b)} $ of the binned {\it chain} (one-sided) a Gibbs measure in the DLR sense (two-sided)? The following theorem gives a positive answer. 

\begin{theo}\label{theo:Gibbs}
The law $\mathbb{P}^{(b)} $ of the binned chain is a Gibbs measure. 
\end{theo}

The proof of this theorem is given in the Appendix. It relies on the well-known Gibbsian character of processes having exponential continuity rate. 

As a conclusion, the influence of past \textit{and} future in estimating the probability to be in the present state decays exponentially with the distance, and there is a unique probability compatible with a given past and future.

\su{Consequences and conclusion} \label{sec:conclusion}

In this paper we have considered rigorously mathematical effects of the binning procedure on the spike train analysis. Especially, we have shown that binning induces naturally long memory effects, even if the initial process is Markovian. 

For a fixed system size, we have excluded possible spurious mathematical consequences of this artificial memory, such as a qualitatively different behavior of the binned chain from the behavior of the original, non-binned, chain. Indeed, when starting with a Markov chain, i.e.\ a process having finite memory, as a model for spike trains, then the binned chain, though of unbounded memory with variable length, will automatically present all good statistical features needed to study its longtime behavior. These good features are the renewal property,  implying factorization of the past, and the exponential decay of the continuity rates.  Here, "renewal property" means that the past can be cut into i.i.d.\ parts of pieces of history in between successive renewal events, implying the ergodic theorem as a simple consequence of the law of large numbers. In other words, for both, binned and original chain, we dispose of a law of large numbers and the convergence to equilibrium will be exponentially fast. Of course, statistical averages of functions will not be the same in the two processes, but their longtime behavior is of the same type. Moreover, although the Markov property is lost here in the one-sided situation, we prove that the Gibbs property -- in the {\it DLR sense} --  remains.




Notice that our proof holds only \textit{when the number of neurons is finite}, and specific singularities in the binned chain, thoroughly analyzed in the context of mathematical statistical physics and coined in terms of {\it phase transitions}, could arise as $N \to \infty .$ Indeed, the exponential decay coefficient $\alpha$ in Proposition \ref{prop:2} is positive. But it depends on $N, $ and we cannot exclude that it converges to $0$ as $N \to +\infty$. In this case, there might exist evidence of first order phase transition, even if the number of neurons is finite, because estimation of probabilities is based on finite rasters: taking a raster of length $T$ with a given number of neurons $N$ and extrapolating statistical properties of the underlying probability for an increasing number of neurons $n_0 < n_1 < \dots \leq N$ could lead to such effects. 

 A {\it second order phase transition}, associated to the notion of critical phenomena, corresponds to the situation where we have a unique  probability compatible with the specification, but where space and time correlations decay algebraically instead of exponentially. A small perturbation on one neuron could in this case trigger long range effects which are power law distributed. Critical phenomena are interesting because they can be classified according to a set of numbers called {\it critical exponents}. Remarkably, critical phenomena observed in nature can be classified into a few "universality classes" sharing the same set of critical exponents, see \cite{hohenberg-halperin:77,ma:76}. There exist various methods to compute critical exponents, the most well known being the renormalization group analysis, \cite{kadanoff:66,wilson:75}. For this reason, researchers are actively seeking evidence of critical phenomena e.g. in the retina \cite{tkacik-schneidman-etal:09}. 
In our case, the binned system cannot exhibit a critical behavior for $N$ finite. This is excluded by classical results on primitive Markov chains and the Perron-Frobenius theorem (spectral gap). However, again, one cannot exclude that binning induces spurious evidence of criticality as extrapolating with a growing number of neurons and we fear that, as $N \to \infty,$ binning could dramatically change the value of the critical exponents, leading to wrong conclusions concerning the universality class. 

Let us finally point out some mathematical directions of situations in which the mathematical consequences of the artificial memory spanned by the binning procedure could be much worse than in the situation described above. Suppose e.g.\ that the law of the original chain is given by a two-sided model exhibiting a phase transition\footnote{Starting by an invariant measure for the two-sided model, not for the one-sided.}, like e.g. the long-range Dyson model with pair potentials that decay polynomially with parameter $1<\alpha<2$ (see Redig and Wang \cite{RW}, van Enter {\it et al.} \cite{ELN}). Then we suspect that the extra memory due to binning added to  long range interactions could give rise to a non-Gibbsian measure as a consequence of the creation of a point of (two-sided) discontinuity. This discontinuity is NOT a critical phenomena but could be fallaciously interpreted as a manifestation of criticality; the discontinuity is just a proof of the fact that the measure is non-Gibbs, it does not correspond to a phase transition of any order. It might even be possible that --  starting from a uniqueness measure of the Dyson-Ising specification -- the binning procedure could yield a lower temperature long-range model for which a hidden phase transition occurs, and this could be misunderstood as the creation of criticality by binning. 

Binning shares similarities with renormalization group transformations for which these types of pathologies (due to scaling transformations \cite{VEFS}) have also been detected, and explained as the manifestation of discontinuities of the renormalized (i.e.\ binned) process. These discontinuities could as well wrongly be interpreted as a critical phenomena. 
In other words, the mathematical question which is interesting in this context is the following. "Can binning induce fallacious evidences of phase transitions?". Heuristically, in statistical physics, a phase transition is observed in a system whose number of degrees of freedom (here: neurons) tends to infinity. On practical grounds, where neurons number is always finite, one proceeds by considering increasing 
sizes and extrapolate to infinity. Efficient methods such as Finite-Size Scaling (see \cite{privman-fisher:84}) allow to nicely extrapolate the properties of the system in the infinite size limit (thermodynamic limit). So, in our case, one has to determine how binning could affect such an extrapolation.

From another point of view the effect of binning has been nicely discussed in a recent paper by   
I. Mastromatteo and M. Marsili \cite{mastromatteo-marsili:11}. They have shown that inference procedures used in statistical mechanics are likely to yield models which are close to a phase transition. Distinguishable models tend to accumulate close to critical points, where the susceptibility diverges in infinite systems in a region where the estimate of inferred parameters is most stable. Their paper suggests that spurious evidences of criticality can be inherent to the way data are considered. Our paper gives a similar warning, in a different context.\\

\paragraph{Ackwnowledgment.} We could like to thanks the referees for helpful and constructive criticism. 

\section{Appendix} 

\begin{proof}{\bf of Proposition \ref{prop:1}}
Fix $ x_{-\infty}^{-1} .$  If $ l(x_{-\infty}^{-1}) = +  \infty ,$ then \eqref{eq:real} is trivially satisfied. Therefore, suppose that $ l(x_{-\infty}^{-1}) = \ell <  \infty .$ Notice that this event is ${\cal F}_{- \ell }^{-1}-$measurable. Let $ A \in {\cal F}_{- \infty}^{- \ell - 1 } .$ Denote by $1_A$ the indicatrix function of event $A$.
Then by definition of $ l(x_{-\infty }^{-1} ) $ and by definition of the binned raster by means of the aggregation map $\pi ,$  
\begin{equation}\label{eq:1}
\E^{(b)}  [1_a ( X_0 ) 1_A  1_{ X_{- \ell }^{-1} = x_{-  \ell }^{-1} } ]  = \E [ 1_{ \pi^{-1} ( a) } ( X_{0}^{\tau - 1})  1_{ \pi^{-1} ( A) } ( X_{-\infty}^{-\ell \tau - 1 } ) 1_{ \pi^{-1} ( x_{-\ell +1}^{-1})  } 1_{ X_{ -\ell \tau }^{(- \ell +1) \tau - 1 }  \equiv z } ] .
\end{equation}
But the RHS can be written as

\begin{eqnarray}\label{eq:10}
&&\E [ 1_{ \pi^{-1} ( a) } ( X_{0}^{\tau - 1})  1_{ \pi^{-1} ( A) } ( X_{-\infty}^{-\ell \tau - 1 } ) 1_{ \pi^{-1} ( x_{-\ell +1}^{-1})  } 1_{ X_{ -\ell \tau }^{(- \ell +1) \tau - 1 }  \equiv z } ]  \nonumber \\
&&= \E \left[ \E \left(  1_{ \pi^{-1} ( a) } ( X_{0}^{\tau -1})  1_{ \pi^{-1} ( x_{-\ell +1}^{-1})  }  | {\cal F}_{- \infty}^{ (- \ell +1) \tau -1 } \right) 1_{\pi^{-1} (A)} 1_{ X_{-  \ell \tau }^{(- \ell +1) \tau -1}  \equiv z } \right]  \nonumber \\
&&= \E \left[ \E \left(  1_{ \pi^{-1} ( a) } ( X_{0}^{\tau -1})  1_{ \pi^{-1} ( x_{-\ell +1}^{-1})  }  | X_{ -\ell \tau }^{(- \ell +1) \tau -1}  \equiv z \right)1_{\pi^{-1} (A)}1_{ X_{ -\ell \tau }^{(-\ell +1) \tau -1}  \equiv z } \right] , 
\end{eqnarray}

where the second line holds trivially because the events $\pi^{-1} ( A ) $ and $\{ X_{ -\ell \tau }^{(- \ell +1) \tau - 1 }  \equiv z \}$ are ${\cal F}_{- \infty}^{ (- \ell +1) \tau -1 }-$measurable, whereas the last line follows from the Markov property of $(X_n )_n$ under $\mathbb{P}$ of order $D$, the fact that $ \tau \geq D$ and the fact that the block $X_{ -\ell \tau }^{(- \ell +1) \tau - 1 }$  
is constrained to be equal to $z$. 

If we define
\begin{equation}\label{eq:transbinned}
 p ( a | x_{- \ell }^{-1}) := \frac{ \E \left(  1_{ \pi^{-1} ( a) } ( X_{0}^{\tau -1})  1_{ \pi^{-1} ( x_{-\ell +1}^{-1})  }  | X_{ -\ell \tau }^{(- \ell +1) \tau -1 }  \equiv z \right) }{\E \left(  1_{ \pi^{-1} ( x_{-\ell +1}^{-1})  }  | X_{ -\ell \tau }^{(- \ell +1) \tau -1 }  \equiv z \right) },
\end{equation}
we obtain, comparing to \eqref{eq:1} and \eqref{eq:10},

\begin{eqnarray*}
&&\E^{(b)} [1_a ( X_0 ) 1_A  1_{ X_{- \ell }^{-1} = x_{-  \ell }^{-1} } ] \\
&&= 
\E \left[ 
p ( a | x_{- \ell }^{-1}) \, \E \left(  1_{ \pi^{-1} ( x_{-\ell +1}^{-1})  }  | X_{ -\ell \tau }^{(- \ell +1) \tau -1 }  \equiv z \right)
1_{\pi^{-1} (A)}1_{ X_{ -\ell \tau }^{(-\ell +1) \tau -1}  \equiv z } \right] 
\\
&&= 
p ( a | x_{- \ell }^{-1}) \, \E \left[ 
 \E \left(  1_{ \pi^{-1} ( x_{-\ell +1}^{-1})  }  | X_{ -\ell \tau }^{(- \ell +1) \tau -1 }  \equiv z \right)
1_{\pi^{-1} (A)}1_{ X_{ -\ell \tau }^{(-\ell +1) \tau -1}  \equiv z } \right] \\
&& = 
p ( a | x_{- \ell }^{-1}) \, \E \left[ 
1_{\pi^{-1} (A)} 1_{ \pi^{-1} ( x_{-\ell +1}^{-1})  } 1_{ X_{ -\ell \tau }^{(-\ell +1) \tau -1}  \equiv z } \right]  ,
\end{eqnarray*}
whence
$$ \E^{(b)} [1_a ( X_0 ) 1_A  1_{ X_{- \ell }^{-1} = x_{-  \ell }^{-1} } ]  =  p ( a | x_{- \ell }^{-1}) \E^{(b)} [1_A  1_{ X_{- \ell }^{-1} = x_{-  \ell }^{-1} } ],$$
and as a consequence, we have shown that  $p ( a | x_{- \ell }^{-1})$ is a regular version of the conditional probability of $ \mathbb{P}^{(b)} \left[ X(0) =a | {\cal F}_{- \infty}^{-1 } \right]  $ on $ \{ X_{- \infty}^{-1} = x_{-\infty}^{- 1 } \} , $ when $  l(x_{-\infty}^{-1}) = \ell  .$  Note that being of variable length, this system of transition kernels is by definition continuous on the set of all semi-infinite past sequences containing at least one $z.$ Indeed, on this set, the transition kernels depend only on a finite portion of the past and are therefore, a fortiori, continuous.  
This concludes our proof. 
\end{proof}

\begin{proof}{\bf of Proposition \ref{prop:2}} 
We suppose first that $ \tau \geq D.$ 
Introduce $Y_n := X_{n \tau }^{(n+1) \tau -1 } , n \in \setZ.$ Since $ \tau \geq D , $  $Y_n $ is a primitive and homogeneous Markov chain of order $1$ under $\mathbb{P},$ and $\mathbb{P}^{(b)} = {\cal L} (  (\pi ( Y_n))_{n \in \setZ} )  $ is the law of the factor chain obtained through the factor map $ \pi $ introduced in \eqref{eq:pi}. Write $q $ for the transition kernel of $Y. $  By our assumptions, $q$ is a strictly positive continuous transition kernel which is locally constant, i.e.\ 
$$ \beta^{q  } ( k) = 0 \mbox{ for all } k \geq 2. $$
Then Theorem 1.1 of \cite{Verbitskiy2011315} can be applied, and it implies that $ \beta^{p^{(b)}} (k) \to 0$ as $k \to \infty, $ without however giving a precise rate of convergence.

In order to obtain a control on the rate of convergence, we rely on the results obtained in \cite{chazottes-ugalde:11}. The notations used there are slightly different from ours, in particular, they work with right infinite sequences of symbols drawn from $\cA $ which represent all possible pasts. If we translate our objects into their framework, then $ \mu := {\cal L} ( X_{-\infty}^0 | \mathbb{P} ) $ is a $D-$step Markov measure in the sense of \cite{chazottes-ugalde:11}. Then Equation (7) of Theorem 3.1 of \cite{chazottes-ugalde:11} implies that there exists a function $ \Phi^{(b)} : \cA_{-\infty}^0 \to \setR $ such that we have convergence 
\begin{equation}\label{eq:potentialbinned}
  \lim_{n \to \infty } \log p^{(b)} ( a | a_{-n}^{-1} ) = \Phi^{(b)} ( a_{-\infty}^0) 
\end{equation}
for all $ a_{- \infty}^0 \in \cA_{-\infty}^0 $ and such that 
\begin{equation}\label{eq:continuitybinned}
\sup_{ a_{-\infty}^0 , b_{-\infty}^0 \in \cA_{-\infty}^0 : a_{-n}^0 = b_{-n}^0 } |  \Phi^{(b)} ( a_{-\infty}^0) - \Phi^{(b)} ( b_{-\infty}^0) | \le C e^{ - \alpha n } , 
\end{equation}
for some $\alpha > 0.$  Due to the uniform continuity of $ ]- \infty  , 0 ] \ni x \mapsto e^x ,$ this implies also that 
$$ \beta^{p^{(b)}} (n ) \le C e^{- \alpha n } , $$
for all $n ,$ and this concludes the proof. 

Finally, if $ \tau < D, $ let $ N = \min \{ k : k \tau \geq D \} $ and set $ \tilde Y_n = X_{n \tau}^{(n + N) \tau - 1 }, n \in \setZ .$ Then the above proof remains true, working with $\tilde Y_n$ instead of $ Y_n.$ 
\end{proof}

\begin{proof}{\bf of Theorem \ref{theo:Gibbs}}
Our proof follows ideas given in \cite{chazottes-ugalde:11}. In \cite{chazottes-ugalde:11},  the authors prove the Gibbs property of a factor chain (i.e.\ of the law of the binned raster) in the sense of "Sinai-Ruelle-Bowen'', see e.g.\ \cite{bowen:98}. One speaks also shortly of "SRB"-Gibbs measures (cf.\ to Definition 2.3 and Theorem 3.1 of \cite{chazottes-ugalde:11}). The SRB-Gibbs property is weaker than the standard DLR-Gibbs property in mathematical statistical mechanics and does not imply that conditioning with respect to the future behaves well. We refer to \cite{Robertoetal} for the hierarchy between the two notions. 

The following proof shows that it is nevertheless possible to use the same approach as the one given in \cite{chazottes-ugalde:11} to prove the Gibbs property also in the DLR sense. 

By Theorem 2.8 of \cite{Robertoetal}, it is sufficient to show the uniform convergence of 
$$ \mathbb{P}^{(b)} \left[ X_0 = a(0)  | X_{-m}^{-1} = a_{-m}^{-1} , X_1^n = a_1^n  \right] , $$
as $m, n \to \infty .$ Under the conditions of our paper, this convergence follows easily from the considerations that we have developed in the proof of Proposition \ref{prop:2}. 

Indeed, let us rewrite 
\begin{equation}
\mathbb{P}^{(b)} \left[ X_0 = a(0)  | X_{-m}^{-1} = a_{-m}^{-1} , X_1^n = a_1^n  \right] = 
\frac{\mathbb{P}^{(b)} \left[  X_{-m}^n = a_{-m}^n  \right]}{\sum_{c \in \cA} \mathbb{P}^{(b)} \left[ X_{-m}^{n} = a_{-m}^{-1}c  a_1^n  \right]} .
\end{equation}
We rewrite the numerator of this expression as 
\begin{multline*}
 \mathbb{P}^{(b)} \left[  X_{-m}^n = a_{-m}^n  \right] = \mathbb{P}^{(b)} \left[  X_{-m}^{-1} = a_{-m}^{-1}  \right] \\
\mathbb{P}^{(b)} \left[  X_0 = a(0) | X_{-m}^{-1} = a_{-m}^{-1}  \right] \mathbb{P}^{(b)} \left[  X_{1}^{n} = a_{1}^{n}  | X_{-m}^0 = a_{-m}^0  \right] .
\end{multline*}
The same kind of expression applies to the denominator. As a consequence, 
\begin{multline}\label{eq:expression1}
 \mathbb{P}^{(b)} \left[ X_0 = a(0)  | X_{-m}^{-1} = a_{-m}^{-1} , X_1^n = a_1^n  \right]  = \\
\frac{\mathbb{P}^{(b)} \left[  X_0 = a(0) | X_{-m}^{-1} = a_{-m}^{-1}  \right]  \mathbb{P}^{(b)} \left[  X_{1}^{n} = a_{1}^{n}  | X_{-m}^0 = a_{-m}^0  \right]}{\sum_{ c \in \cA} \mathbb{P}^{(b)} \left[  X_0 = c | X_{-m}^{-1} = a_{-m}^{-1}  \right]  \mathbb{P}^{(b)} \left[  X_{1}^{n} = a_{1}^{n}  | X_{-m}^0 = a_{-m}^{-1} c   \right] } .
\end{multline}
But by \eqref{eq:potentialbinned}, and following \cite{chazottes-ugalde:11},  we have uniform convergence of 
$$ \mathbb{P}^{(b)} \left[  X_0 = a(0) | X_{-m}^{-1} = a_{-m}^{-1}  \right] \to e^{ \Phi^{(b)} ( a_{-\infty}^0)} $$
and of 
$$ \mathbb{P}^{(b)} \left[  X_0 = c | X_{-m}^{-1} = a_{-m}^{-1}  \right]  \to e^{ \Phi^{(b)} ( a_{-\infty}^{-1} c) } ,$$
as $ m \to \infty .$ 

Moreover, 
$$ \mathbb{P}^{(b)} \left[  X_{1}^{n} = a_{1}^{n}  | X_{-m}^0 = a_{-m}^0  \right] = \prod_{k=1}^n \mathbb{P}^{(b)} \left[  X_k = a (k)   | X_{-m}^{k-1} = a_{-m}^{k-1}  \right] $$
such that we can rewrite \eqref{eq:expression1} as
\begin{multline}\label{eq:expression2}
 \mathbb{P}^{(b)} \left[ X_0 = a(0)  | X_{-m}^{-1} = a_{-m}^{-1} , X_1^n = a_1^n  \right]  = 
\mathbb{P}^{(b)} \left[  X_0 = a(0) | X_{-m}^{-1} = a_{-m}^{-1}  \right] \cdot \\
\left[ \sum_{c \in \cA} \mathbb{P}^{(b)} \left[  X_0 = c | X_{-m}^{-1} = a_{-m}^{-1}  \right] \prod_{k=1}^n \left( \frac{\mathbb{P}^{(b)} \left[  X_k = b(k)   | X_{-m}^{k-1} = b_{-m}^{k-1}  \right]}{\mathbb{P}^{(b)} \left[  X_k = a (k)   | X_{-m}^{k-1} = a_{-m}^{k-1}  \right] } \right) \right]^{-1} 
\end{multline}
where $b(k) = a (k) $ for all $ k \neq 0, $ $b(0) = c .$ It remains to show that 
$$ \prod_{k=1}^n \left( \frac{\mathbb{P}^{(b)} \left[  X_k = b(k)   | X_{-m}^{k-1} = b_{-m}^{k-1}  \right]}{\mathbb{P}^{(b)} \left[  X_k = a (k)   | X_{-m}^{k-1} = a_{-m}^{k-1}  \right] } \right) $$
converges as $ m, n \to \infty .$ For any fixed $k, $ we have, still by \eqref{eq:potentialbinned}, that 
$$ \frac{\mathbb{P}^{(b)} \left[  X_k = b(k)   | X_{-m}^{k-1} = b_{-m}^{k-1}  \right]}{\mathbb{P}^{(b)} \left[  X_k = a (k)   | X_{-m}^{k-1} = a_{-m}^{k-1}  \right] }  \to e^{ \Phi^{(b)} ( a_{-\infty}^k ) - \Phi^{(b)} ( b_{-\infty}^k ) }  \; \mbox{ as } m \to \infty .$$ 
Moreover, we have that 
\begin{equation}\label{eq:convdomine}
 \frac{\mathbb{P}^{(b)} \left[  X_k = b(k)   | X_{-m}^{k-1} = b_{-m}^{k-1}  \right]}{\mathbb{P}^{(b)} \left[  X_k = a (k)   | X_{-m}^{k-1} = a_{-m}^{k-1}  \right] }  \le e^{ C e^{ - \alpha k } } , 
\end{equation}
which follows from the representation 
$$   \mathbb{P}^{(b)} \left[  X_k = b(k)   | X_{-m}^{k-1} = b_{-m}^{k-1}  \right] = \int_\Omega p^{(b)} ( b(k) | b^{k-1}_{-m } \omega_{-\infty}^{-m-1} ) \mathbb{P}^{(b)} (d\omega ) p^{(b)} ( b(k) | b^{k-1}_{-m } \omega_{-\infty}^{-m-1} ) ,$$
from \eqref{eq:potentialbinned} and from the fact that $ b^k_1 = a^k_1.$ 

Now, \eqref{eq:convdomine} implies by Lebesgue's theorem of dominated convergence that 
$$ \sum_{k=1}^n \log \left( \mathbb{P}^{(b)} \left[  X_k = b(k)   | X_{-m}^{k-1} = b_{-m}^{k-1}  \right] \right) - 
\log \left( \mathbb{P}^{(b)} \left[  X_k = a(k)   | X_{-m}^{k-1} = a_{-m}^{k-1}  \right] \right) $$
converges, as $ n \to \infty ,$ and this concludes our proof.  
\end{proof}

\bibliographystyle{plainnat}
\bibliography{biblio}

\end{document}